\documentclass[%
 aip,
 amsmath,amssymb,
 reprint,%
]{revtex4-1}

\usepackage{graphicx}% Include figure files
\usepackage{dcolumn}% Align table columns on decimal point
\usepackage{bm}% bold math
\usepackage[utf8]{inputenc}
\usepackage[T1]{fontenc}
\usepackage{mathptmx}
\usepackage{etoolbox}

\usepackage{amssymb}
\usepackage[version=4]{mhchem}
\usepackage{color}
\usepackage[colorlinks=true, letterpaper=true, pdfstartview=FitV, linkcolor=blue, citecolor=blue, urlcolor=blue]{hyperref}
\usepackage{lineno}
\makeatletter
\def\@email#1#2{%
 \endgroup
 \patchcmd{\titleblock@produce}
  {\frontmatter@RRAPformat}
  {\frontmatter@RRAPformat{\produce@RRAP{*#1\href{mailto:#2}{#2}}}\frontmatter@RRAPformat}
  {}{}
}%
\makeatother
\begin{document}

\preprint{AIP/123-QED}

\title[]{Hexagonal InOI monolayer: a 2D phase-change material combining topological insulator states and piezoelectricity}
% Force line breaks with \\
\author{Wenhui Wan}
\email{wwh@ysu.edu.cn}
\thanks{corresponding authors}
\author{Xinyue Liu}
\author{Yanfeng Ge}
\author{Ziqang Li}
\author{Yong Liu}
%\email{yongliu@ysu.edu.cn}
%\thanks{These authors contributed equally as corresponding authors}
\affiliation{State Key Laboratory of Metastable Materials Science and Technology $\&$ Hebei Key Laboratory of Microstructural Material Physics, School of Science, Yanshan University, Qinhuangdao 066004, P. R. China.}
\date{\today}% It is always \today, today,
             
\begin{abstract}
Two-dimensional (2D) phase-change materials (PCMs) with moderate transition barriers and distinctly contrasting properties are highly desirable for multifunctional devices, yet such systems remain scarce. Using first-principles calculations, we propose a hexagonal InOI monolayer as a promising 2D PCM. This material exhibits two distinct polymorphs: an energetically favorable T$^{\prime}$ phase and a metastable T phase, differentiated by iodine atom positions. The T$^{\prime}$-to-T structural phase transition features a moderate energy barrier $E_b$ of 72.1 meV per formula unit, facilitating reversible switching. Notably, strain engineering tailors the electronic transition, inducing either a metal-to-topological-insulator or a metal-to-normal-insulator transformation. Additionally, this phase transition modulates the piezoelectric response and shifts optical absorption from the infrared to the visible range. These multifunctional properties make 2D hexagonal InOI highly promising for applications in non-volatile memory, low-contact-resistance spintronics, and optical switching devices.
\end{abstract}

\maketitle

2D phase-change materials (PCMs) exhibit tunable electronic and optical properties through  reversible phase transitions,\cite{Kim2023} with wide potential applications in low-resistance electronic contacts, nonvolatile memory, photodetectors, and optical energy harvesting.\cite{Izadi2025} The current experimental strategies for driving the structural phase transition include the use of defects, ion intercalation, alloying, temperature, strain, and electric fields.\cite{Gao2025} Despite significant advances in 2D PCMs, several critical challenges remain, such as high transition energy barriers, inefficient charge and heat transport, mechanical brittleness, and metastability issues after phase transitions.\cite{Ghasemi2020,Chen2022}

Transition metal dichalcogenides (TMDs) MX$_2$ (M = Mo, W; X = S, Se, Te) are extensively studied 2D PCMs. Their monolayers typically exhibit three polymorphs: a semiconducting 1H phase, a metallic 1T phase, and a semimetallic 1T$^{\prime}$ phase, where the latter two phases host unique magnetic, superconducting, and topological properties.\cite{Sokolikova2020} The 1H-to-1T$^{\prime}$ phase transition can be induced through defects,\cite{Xiao2024} ion intercalation,\cite{Wang2021}  strain,\cite{Sabbaghi2024} and electrical gating.\cite{Eshete2023} Despite the electronic contrast between two phases, their switching performance in practical applications is limited by a substantial energy barrier of 0.69–1.51 eV/formula unit (f.u.).\cite{Demirkol2022}
%a substantial energy difference (0.04 - 0.55 eV/formula unit (f.u.)) \cite{Duerloo2014} and 

Group-III monochalcogenides MX (M = Ga, In; X = S, Se, Te) represent a distinct class of 2D PCMs, which consist of a quadruple layer with an X–M–M–X stacking sequence.~\cite{Cai2019} Their monolayers stabilize in two semiconducting phases: 1H and 1T.\cite{Lee2025,Gutierrez2024,Yu2024} The ground-state 1H phase ($D_{3h}$ symmetry) maintains mirror symmetry with vertically aligned chalcogen atoms, whereas the 1T phase ($D_{3d}$ symmetry) exhibits a 180$^{\circ}$ rotation between chalcogen layers. The inversion symmetry in the 1T phase suppresses nonlinear optical responses such as second-harmonic generation.\cite{Nitta2020} These two phases display a small energy difference (13–20 meV/f.u.)\cite{Zolyomi2014,Aliyev2025} and a moderate switching barrier (0.15–1.10 eV/f.u.).\cite{Sun2020,Nitta2020,Lee2025} However, both phases feature similar electronic structures with a sombrero-shaped valence band dispersion.\cite{Zolyomi2014,Zhou2016} The 1H-to-1T phase transition exhibits minimal electronic contrast, which limits device functionality.

In this work, motivated by recent experimental realization of 2D hexagonal InO crystal \cite{KakanakovaGeorgieva2021} and surface halogenation techniques, \cite{Tang2021} we propose an InOI monolayer (ML) as a 2D PCM featuring moderate switching barriers (72 meV/f.u.) and starkly contrasting physical properties. First-principles calculations reveal that while InO$X$ ($X$ = Cl, Br, I) MLs can be derived from the presynthesized InO ML, only the InOI ML exhibits reversible PCM behavior. Its structural phase transition undergoes a strain-modulated metal-to-insulator transformation. Moreover, InOI ML shows phase-dependent topological state, optical absorption, and piezoelectric response. These results indicate the InOI ML as a promising 2D PCM with integrated topological and piezoelectric functionalities.

\begin{table*}[tb]
	%	\small
	\caption{The lattice constant ($a$), bond lengths ($d_{\rm In-O}$, $d_{{\rm In}-X}$), band gap ($E_{g}$) calculated by HSE06+SOC method, elastic constants ($C_{11}$ and $C_{12}$), Young's modulus ($Y$), Poisson's ratio ($\nu$), 2D polarization $P_z$ (the dipole moment per area), and  piezoelectric coefficients ($e_{22}$, $e_{32}$, $d_{22}$, and $d_{32}$) of InO$X$ ($X$ = Cl, Br, I) MLs.}
	\label{table1}
	\begin{ruledtabular}
		\centering % 使表格居中  
		\begin{tabular*}{1\textwidth}{@{\extracolsep{\fill}}*{15}{c}}
			& phase &$a$  &$d_{\rm In-O}$  &$d_{{\rm In}-X}$  &$E_{g}$  &$C_{11}$  &$C_{12}$  &$Y$ & $\nu$ &$P_z$    &$e_{22}$  &$e_{32}$  &$d_{22}$  &$d_{32}$ \\\cline{3-5} \cline{7-9} \cline{12-13} \cline{14-15} 
			
			& &\multicolumn{3}{c}{(\AA)} &(eV)     &\multicolumn{3}{c}{(N/m)}   &  & (pC/m)     &\multicolumn{2}{c}{(pC/m)}     &\multicolumn{2}{c}{(pm/V)}       \\
			\hline
			InOCl &T&3.508&2.158&2.700&3.774 &79.88&33.71&65.65&0.422&$-1.56$&367.76 &$-21.92$ &7.97  &$-0.20$ \\
			InOBr &T&3.567&2.181&2.834&2.207 &79.78&31.69&67.20&0.397&$-0.47$&372.64&$-5.78$ &7.75  &$-0.05$   \\
			InOI  &T&3.677&2.218&3.063&0.079 &66.24&28.71&53.80&0.433&$-3.46$&843.74&33.26 &22.48 &0.35  \\
			InOI  &T$^{\prime}$&3.667&2.186&2.808&0.000 &66.28 &36.55&46.12&0.551&0.00&        &      &       &        \\
		\end{tabular*}
	\end{ruledtabular}
\end{table*}
%\section{Computational method}
First-principles calculations were performed with the Vienna Ab initio Simulation Package (VASP),~\cite{q28} using the projector augmented wave (PAW)~\cite{q29} pseudopotentials and the Perdew, Burke, and Ernzerhof (PBE)~\cite{q30} exchange-correlation functionals. A 15 \AA\ vacuum layer was included to prevent interactions between periodic images. We employed a kinetic energy cutoff of 550 eV and a $12 \times 12 \times1$ k-point mesh~\cite{q31} for Brillouin zone (BZ) sampling. Convergence criteria were set to 10$^{-6}$ eV for total energy and 10$^{-3}$ eV/\AA\ for atomic force. Band structures were calculated using the Heyd-Scuseria-Ernzerhof (HSE06) screened hybrid functional,~\cite{q32} including the spin-orbit coupling (SOC) effect. The topologically protected edge states and Wannier charge centers (WCC) were calculated using the WannierTools code.~\cite{Wu2018} Piezoelectric coefficients were derived from the density functional perturbation theory.\cite{Baroni1986} Phonon dispersion relations were obtained by the finite displacement method as implemented in the Phonopy code,\cite{q33} using a $4\times4\times1$ supercell. Molecular dynamics (MD) simulations were conducted in the NVT ensemble on a $4\times4\times1$ supercell for 7 ps.

%\section{Results and discussion}
%\subsection{The formation of hexagonal InOX monolayer}

\begin{figure}[tb]
	\centering
	\includegraphics[width=0.45\textwidth]{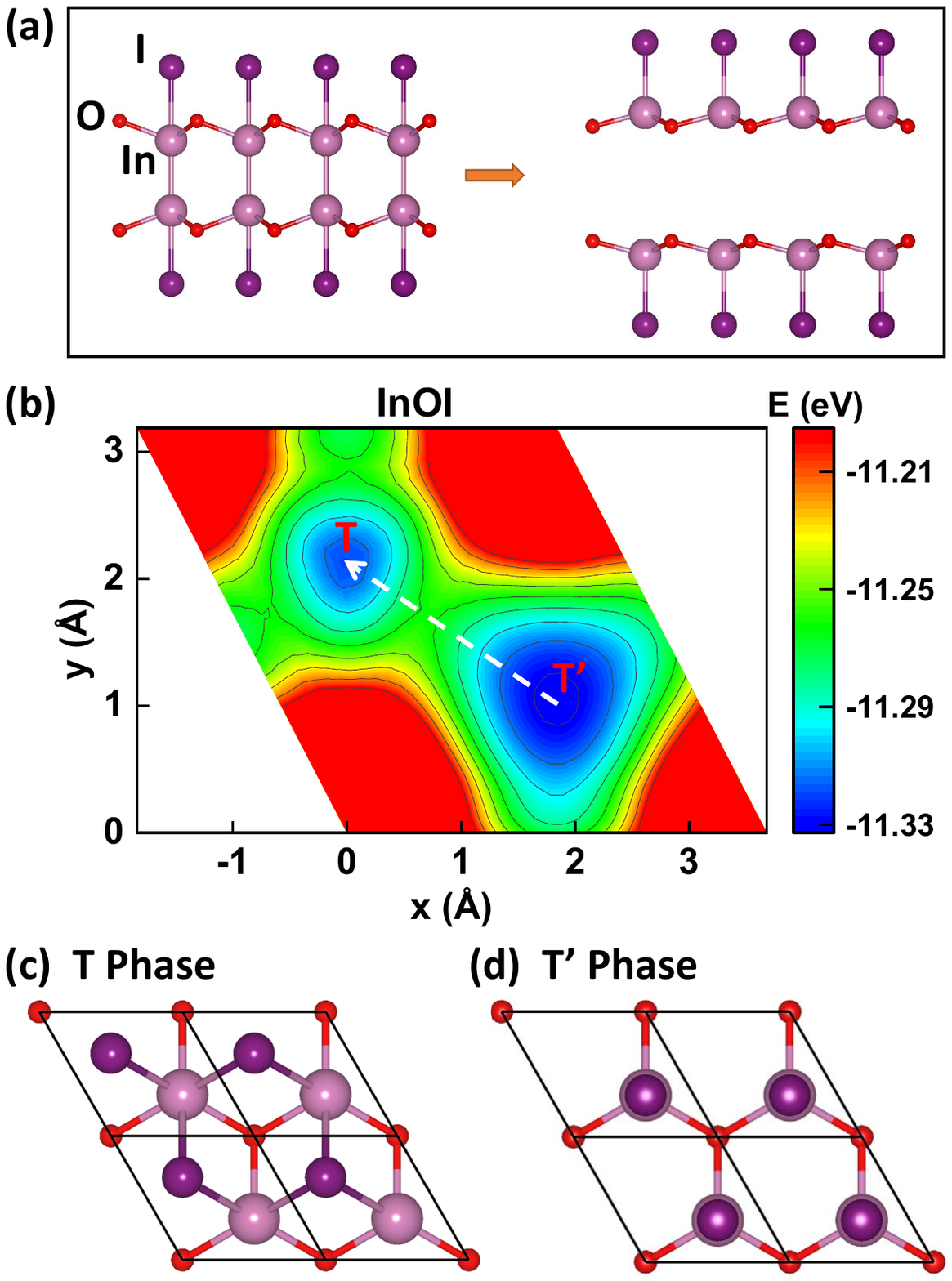}
	\caption{(a) Surface iodination decomposes the hexagonal lattice of InO ML into two InOI MLs. %The In, O, and I atoms are presented by the ,red, and purple balls. 
		(b) The energy of InOI ML as a function of Iodine position. The white arrow represents the phase transition path from the T$^{\prime}$ to the  T phases. (c) and (d) are the lattice structures of the T and T$^{\prime}$ phases of InOI ML, respectively.}\label{wh1}
\end{figure} 

First-principles calculations reveal that surface halogenation spontaneously cleaves In-In bonds in presynthesized hexagonal InO ML [Fig.~\ref{wh1}(a)], forming Janus InO$X$ ($X$ = Cl, Br, I) MLs. The enthalpy change ($\bigtriangleup H$) for this decomposition is estimated as 
\begin{eqnarray}
	\bigtriangleup H = E_{{\rm{InO}}X}-E_{\rm{InO}}-E_{X},
\end{eqnarray}
where $E_{{\rm{InO}}X}$, $E_{\rm{InO}}$, and $E_{X}$ are the energies of InO$X$ ML, InO ML, and a single halogen atom $X$ in its stable gas phase or solid phase, respectively. The calculated $\bigtriangleup H$ is $-1.606$, $-1.035$, and $-0.119$ eV/f.u. for InOCl, InOBr, and InOI MLs, respectively, confirming that this synthesis pathway is exothermic and feasible.

To determine the stable structures of InO$X$ ($X$ = Cl, Br, I) MLs, we computed the total energy as a function of halogen position. The potential energy surface of InOI ML [Fig.~\ref{wh1}(b)] exhibits two local minima: above the hexagonal hollow site [T phase, Fig.~\ref{wh1}(c)] and the top of the In atom [T$^{\prime}$ phase, Fig.~\ref{wh1}(d)]. The T$^{\prime}$ phase is energetically more stable than T phase for InOI ML, while InOCl and InOBr MLs exclusively stabilize in the T phase [Fig. S1(a) and S1(b)]. Compared with Cl and Br, the smaller electronegativity difference between Indium and Iodine enables competition between ionic and covalent bonding, leading to structural polymorphism in InOI ML. The phonon dispersion relation of these four hexagonal InO$X$ MLs, including T-InOCl, T-InOBr, T-InOI, and T$^{\prime}$-InOI MLs, show no imaginary frequencies, confirming their structural stability [Fig. S2]. 

Table~\ref{table1} lists the optimized geometric parameters of InO$X$ ($X$ = Cl, Br, I) MLs. In the T phase, lattice constants ($a$), In-O bond lengths ($d_{\rm In-O}$), and In-$X$ bond lengths ($d_{{\rm In-}X}$) increase with the ionic radius of the halogen atom $X$. Notably, T$^{\prime}$-InOI ML exhibits reduced bond lengths ($d_{\rm In-O}$ and $d_{\rm In-I}$) compared to the T phase, indicating an enhanced In-O and In-I bonding. 

\begin{figure}[tb]
	\centering
	\includegraphics[width=0.45\textwidth]{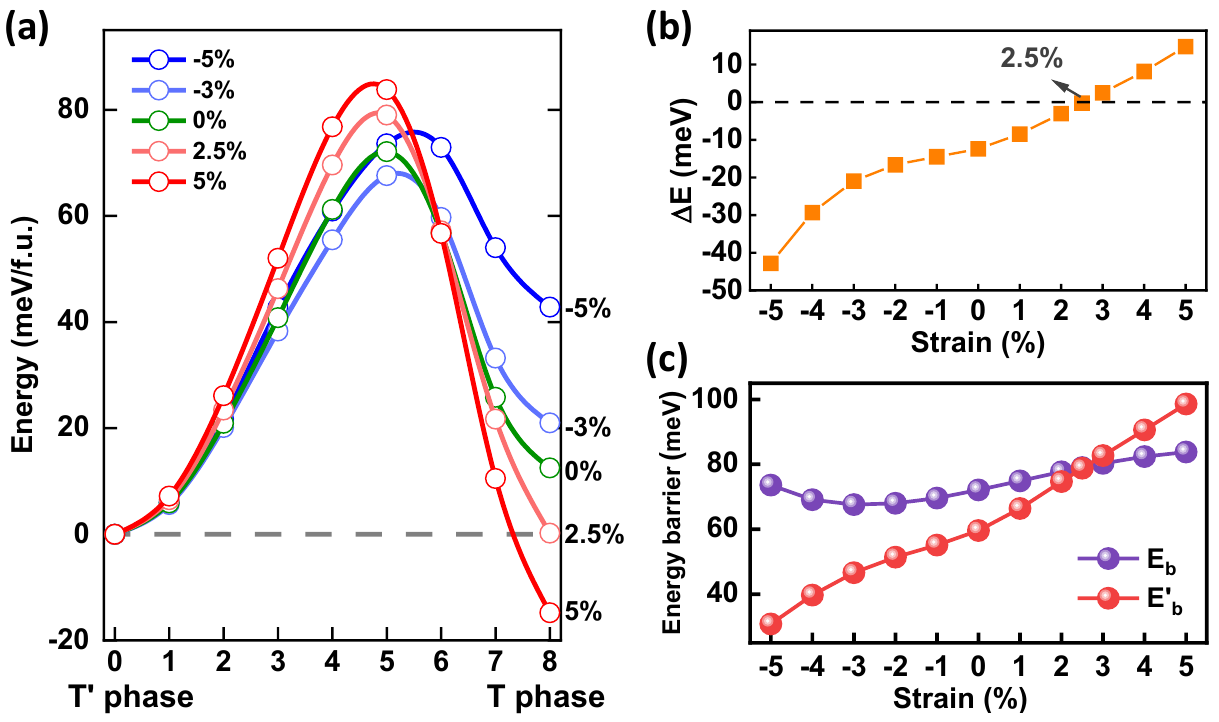}
	\caption{(a) The energy profile along the transition path from the T$^{\prime}$ phase to T phase through the diffusion of I atom. (b) The energies of T$^{\prime}$ phase relative to T phase at different biaxial strains. (c) Energy barrier ($E_b$) of T$^{\prime}$-to-T phase transition and that ($E'_b$) of reverse T-to-T$^{\prime}$ phase transition as a function of biaxial strains.}\label{wh2}
\end{figure}

We investigated the phase-change properties of InOI ML, whose T and T$^{\prime}$ phases exhibit nearly identical lattice constants (0.27\% mismatch) and a small energy difference (11 meV/f.u.). The T$^{\prime}$-to-T phase transition proceeds through the iodine diffusion, as labeled in Fig.~\ref{wh1}(b). The nudged elastic band (NEB) calculation\cite{Henkelman2000} reveals a moderate energy barrier $E_b$ of 72.1 meV/f.u. at the equilibrium lattice constant of T$^{\prime}$-InOI ML [Fig.~\ref{wh2}(a)]. The energy barrier $E_b'$ for the inverse transition (T-to-T$^{\prime}$) is 59.6 meV/f.u., which ensures the stability of the metastable T-InOI ML.

Strain effectively modulates this structural phase transition. We defined the biaxial strain as $\varepsilon = (a/a_0-1)\times 100\%$, where $a$ and $a_0$ are the lattice constants with and without strain, respectively. Tensile strain maintains T$^{\prime}$ phase stability up to 2.5\%. At greater tensions, the T phase becomes energetically favorable over T$^{\prime}$ phase [Fig.~\ref{wh2}(b)]. The energy barrier $E_b$ for T-to-T$^{\prime}$ transition continuously rises to 83.8 meV/f.u. at 5\% tensile strain [Fig.~\ref{wh2}(c)]. Conversely, compressive strain further stabilizes the T$^{\prime}$ phase. Energy barrier $E_b$ lowers to 67.4 meV/f.u. at $-3$\% strain before rising at larger compressions. 
On the other hand, the energy barrier $E_b'$ for the inverse transition (T-to-T$^{\prime}$) increases and decreases under tensile and compressive strains, respectively [Fig.~\ref{wh2}(c)].
Notably, both $E_b$ and $E_b'$ are much lower than that of the 1H-to-1T$^{\prime}$ phase transition of TMDs (0.69–1.51 eV/f.u.),\cite{Demirkol2022} indicating superior phase-switching performance in InOI ML.

\begin{figure}[tb]
	\centering
	\includegraphics[width=0.45\textwidth]{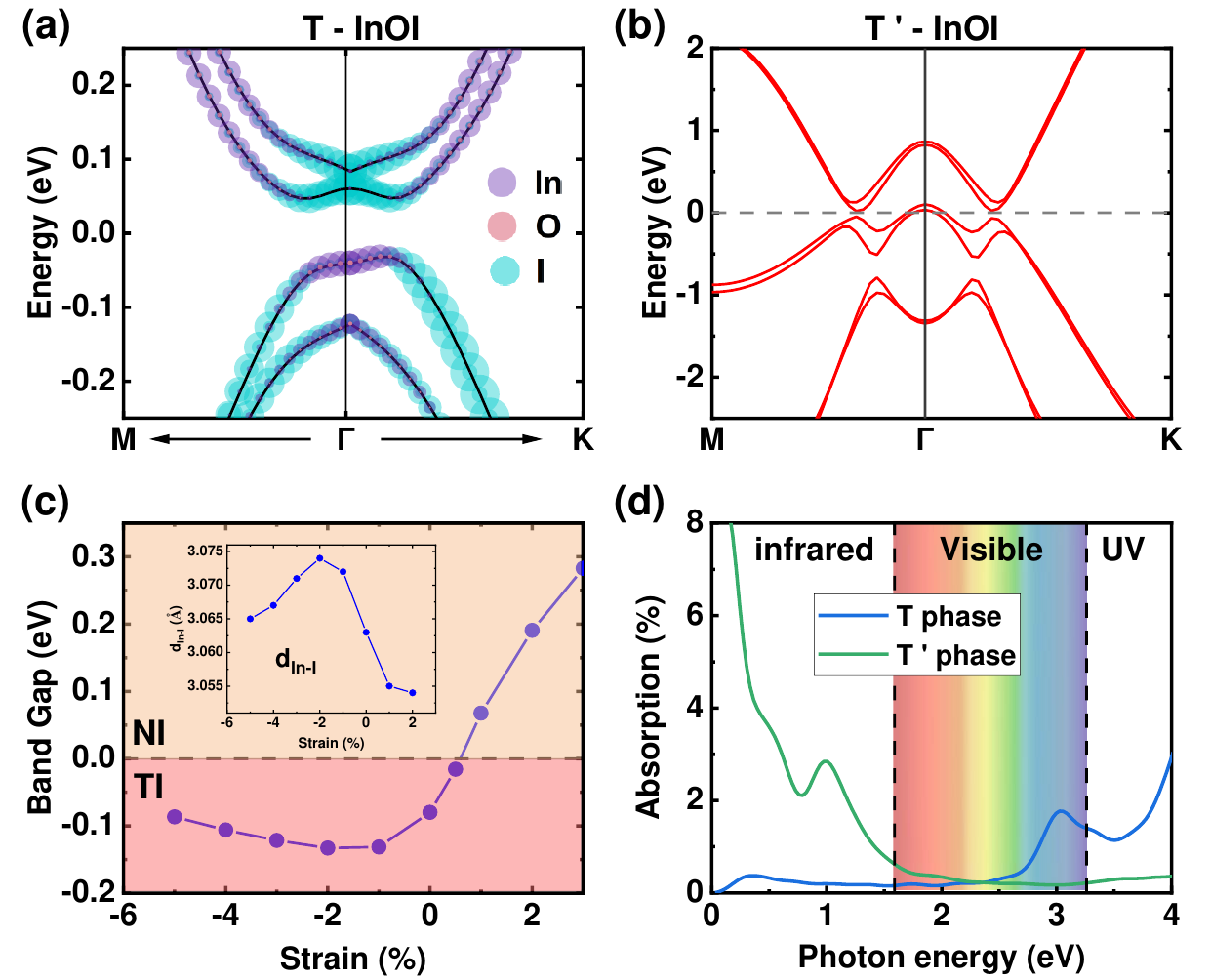}
	\caption{(a) The band inversion at the $\Gamma$ point for T-InOI ML. (b) The band structure of T$^{\prime}$-InOI ML. (c) The band gap of T-InOI ML as a function of biaxial strains. The positive and negative band gaps respond to normal insulator (NI) and topological insulator (TI), respectively. (d) Absorption spectra of T and T$^{\prime}$ phases of InOI ML.}\label{wh3}
\end{figure}

We demonstrated that the T$^{\prime}$-to-T structural phase transition in InOI ML induces a metal-to-topological-insulator transformation. The band structures of T and T' phases of InOI ML exhibit band splitting due to Rashba-type SOC in the asymmetric structure [Figs. S3(c) and S3(d)]. The T-InOI ML exhibits a band gap of 0.079 eV with a band inversion at the $\Gamma$ point [Fig.~\ref{wh3}(a)], characterized by orbital switching between the In-$s$-dominated conduction band minimum (CBM) and I-$p_{x(y)}$-dominated valence band maximum (VBM) [Fig. S4(a)]. To confirm the emergence of nontrivial topology of T-InOI ML, Fig. S5(a) displays helical edge states forming linearly dispersing bands along the $x$ axis edge within the bulk gap; Fig. S5(b) presents the Wannier charge center (WCC) evolution: the evolution curves intersect the arbitrary reference lines once, yielding a topological invariant Z$_2$ = 1. These findings establish T-InOI ML as a promising topological insulator (TI) for dissipationless spintronics. In contrast, T$^{\prime}$-InOI ML exhibits metallic behavior [Fig.~\ref{wh3}(b)]. Though the SOC effect opens mini-gaps at band crossings, no global band gap emerges throughout the BZ. The electron states near the Fermi level are dominated by the In-$s$, I-$p_x$, and I-$p_y$ orbitals [Fig. S6(a)]. Double band inversions occur at the $\Gamma$ point [Fig. S6(b)], which makes T$^{\prime}$-InOI a normal metal. Future experiments could fabricate lateral homojunctions integrating the topological insulator T-InOI ML with a metallic T$^{\prime}$-InOI ML for spintronic devices with low-resistance contacts.~\cite{Han2021}
 
Biaxial strains can modulate the electronic phase transformation in InOI ML. T$^{\prime}$ phase maintains its metallic character across -5\% to 5\% strain ranges. T phase, however, exhibits strain sensitivity. Tensile strains exceeding 0.5\% convert it from a TI to a normal insulator (NI) through band gap closure and reopening [Figs.~\ref{wh3}(c) and S5(c)]. Under tensile strains, the T$^{\prime}$-to-T structural phase transition causes a metal-to-NI transformation. Conversely, compressive strains preserve the TI state while tuning the band gap of T-InOI ML. Under compressive strains from -1\% to -5\%, the fundamental band gap exhibits a nonmonotonic behavior, first increasing to a maximum at -2\% strain and then decreasing. This trend is attributed to the competition between an increasing direct gap at $\Gamma$ and enhanced band splitting near $\Gamma$ [Figs. S5(d-h)]. The maximum band gap reaches 133 meV at -2\% strain, which is close to the band gap of 1T$^{\prime}$-\ce{WTe2} (141 meV).\cite{Zheng2016} 

In the T-InOI ML, the In-$s$ orbital hybridizes with the O-$p$ orbitals. The resulting anti-bonding state $\phi^*$ further hybridizes with I-$p$ orbitals, forming the anti-bonding state $\psi^*$ (primarily contributed by In-$s$ state) in the CBM and the bonding state $\psi$ (primarily I-$p$) in the VBM [Fig. S4(b) and S4(c)]. Tensile strain increases the In-O bond length, thereby lowering the energy of $\phi^*$. However, the In-I bond length ($d_{\rm In-I}$) exhibit an anomalous decrease under tensile strain, alongside an increase in the in-plane cosine of In-I bonds. This enhances the In-I hybridization and increases the band gap between $\psi^*$ and $\psi$. Conversely, compressive strain reduces the bandgap and induces band inversion under SOC. 
The Bader charge analysis [Fig. S7] shows that the non-monotonic In$\rightarrow$I charge transfer leads to the initial lengthening and subsequent shortening of the In–I bond beyond -2\% strain [inset of Fig.~\ref{wh3}(c)]. This nonlinear change in bond length causes the band gap's irregular variation under compressive strain.

Combined the results in Fig.~\ref{wh2}(c) and \ref{wh3}(c), moderate compressive strains ($-3$\% to 0\%) simultaneously reduce the T$^{\prime}$-to-T energy barriers, while enhancing nontrivial topological properties in T-InOI ML. The dual tunability of both band topology and structural phase transition through strain engineering establishes InOI ML as a platform for controllable topological electronic devices.

Beyond electronic structure modifications, the T$^{\prime}$-to-T phase transition alters the optical properties of InOI ML. We calculated the normalized optical absorbance~\cite{q36} of InOI ML. Besides interband transitions, the intraband contributions to the permittivity in the metallic T$^{\prime}$ phase were described by the Drude model for both its real and imaginary parts:
\begin{eqnarray}
	Re[\varepsilon^{intra} (\omega)] &=1-\omega_p^2/(\omega^2+\gamma^2 ),\\
	Im[\varepsilon^{intra} (\omega)] &=\gamma\omega_p^2/(\omega^3+\omega\gamma^2).
\end{eqnarray}
Here, $\gamma$ represents a lifetime broadening, while $\omega_p$ is the free electron plasma frequency tensor. We used a lifetime broadening $\gamma$ of 0.018 eV according to previous work\cite{Laref2013} and got $\omega_p$ from the first-principles calculations.\cite{Setten2009}
The absorption spectra reveal distinct characteristics: the metallic T$^{\prime}$ phase shows strong infrared absorption, while the semiconducting T phase exhibits prominent visible-range absorption [Fig.~\ref{wh3}(d)]. This sharp contrast in optical response suggests potential applications of InOI ML in optical switches and optical memories.

In addition, the T and T$^{\prime}$ phases exhibit distinct piezoelectric properties. The T phase of the InOI ML exhibits an out-of-plane polarization originating from its Janus structure. The electronegativity differences between In and O or I create two oppositely oriented dipole moments, directed from the O and I layers toward the central In layer [Fig. S8(a)]. 
With the Berry phase method,\cite{KingSmith1993} we calculated the net dipole moment per unit cell $D$ and 2D polarization $P_z$ of T-InOI ML as $-0.025$ $e\cdot$\AA\ and $-3.46$ pC/m, respectively, which is along the $-z$ direction. 
This direction is further verified by the plane-averaged electrostatic potential along the $z$ axis [Fig. S8(b)], which shows a higher vacuum level on the I side than on the O side, consistent with the direction of net polarization $P_z$ and built-in electric field. Meanwhile, polarization $P_z$ produces a polarization-induced electric field outside the T-InOI ML. In comparison, metallic T$^{\prime}$-InOI ML exhibits no polarization field due to screening by itinerant electrons. Therefore, the T$^{\prime}$-to-T phase transition offers a phase-dependent electric field to modulate the physical properties of adjacent 2D materials.

In addition to the intrinsic polarization $P_z$, the T-InOI ML exhibits piezoelectric responses in both in-plane and out-of-plane directions. T-InOI ML has the $C_{3v}$ point group symmetry and the mirror symmetry about the $xz$ plane [Fig.~\ref{wh1}(c)]. The independent piezoelectric stress coefficients are $e_{22}$ and $e_{32}$, which describe how in-plane polarization $P_y$ and out-of-plane polarization $P_z$ vary with in-plane uniaxial strain $\varepsilon_{2}$ along the $y$ axis, respectively. The piezoelectric strain coefficients $d_{22}$ and $d_{32}$ can be obtained as
\begin{equation}
	d_{22}= \frac{e_{22}} {C_{22}-C_{12}}, \qquad
	d_{32}= \frac{e_{32}} {C_{22}+C_{12}}.
	\label{e3.31}
\end{equation}
Here, $C_{22}$ and $C_{12}$ are elastic constants. For a hexagonal lattice, $C_{22}$ is equal to $C_{11}$. The calculated elastic constants and piezoelectric coefficients are summarized in Table~\ref{table1}. Due to its small elastic constants, T-InOI ML has high $d_{22}$ (22.48 pm/V) and $d_{32}$ (0.35 pm/V), which exceed those of 2D Janus group-III chalcogenide ($d_{22} = 1.91 - 8.47$ pm/V; $d_{32} = 0.07 - 0.46$ pm/V)~\cite{q13} and 2D Janus metal dichalcogenides ($d_{22} = 2.26 - 5.30$ pm/V; $d_{32} = 0.007 - 0.03$ pm/V).~\cite{q24} Therefore, T-InOI ML has potential piezoelectric applications, a property absent in metallic T$^{\prime}$ phase. 

Mechanical rigidity is a critical issue for PCEs. We calculated the Young's modulus $Y=(C_{11}^2-C_{12}^2)/C_{11}$ and Poisson's ratio $\nu=C_{12}/C_{11}$ for the hexagonal InOI ML. As shown in Table~\ref{table1}, the $Y$ of T- or T$^{\prime}$-InOI MLs are lower than those of graphene (345 N/m)~\cite{Kudin2001} and \ce{MoS_2} (118 N/m)~\cite{q43}, demonstrating their flexibility. Furthermore, their Poisson's ratio $(\nu > 1/3)$ indicates ductile behavior according to the Frantsevich rule,\cite{Frantsevich1982} making InOI ML suitable for PCM applications.

Previous works have demonstrated other ways to drive the structural phase transition except for strains.\cite{Yang2023,He2022,Wang2017} Structural relaxation calculations show that carrier doping in the InOI monolayer (–0.3 to 0.3 $e$/f.u.) mimics strain: electron (hole) doping increases (decreases) the lattice constant [Fig. S9(a)]. Hole doping stabilizes the T$^{\prime}$ phase and raises the T$^{\prime}$-to-T transition barrier $E_b$, while electron doping reduces both the T–T$^{\prime}$ energy difference and $E_b$ [Fig. S9(c)]. Defect calculations in $4 \times 4 \times 1$ supercells show the T phase is stabilized only by I vacancies ($V_{\rm I}$), whereas O vacancies($V_{\rm O}$) and In vacancies ($V_{\rm In}$), as well as the intrinsic structure, favor the T$^{\prime}$ phase [Fig. S9(d)]. Introducing I vacancies is thus proposed to drive the T$^{\prime}$-to-T transition for accessing its topological insulator state.
Furthermore, an electric field can also induce the structural phase transition. The critical electric field strength $E_c$ to switch the T$^{\prime}$-to-T phase transition is estimated by $E_c=\bigtriangleup H/\bigtriangleup D$,\cite{Hu2023} where $\bigtriangleup D$ is the difference of the electric dipole moment between the T$^{\prime}$ and T phases, whose value is 0.025 $e\cdot$\AA; the $\bigtriangleup H$ is the enthalpy difference that is approximated as the energy difference $\bigtriangleup E = 11$ meV. The calculated $E_c$ is estimated to be 0.44 V/\AA.

%\begin{figure}[tb]
%	\centering
%	\includegraphics[width=0.45\textwidth]{fig4.pdf}
%	\caption{(a) The electron mobility of InO$X$ MLs. (b) The hole mobility of InO$X$ MLs.}\label{wh4}
%\end{figure}
Compared to InOI ML, InOCl and InOBr MLs have larger band gaps of 3.774 eV and 2.207 eV, respectively [Fig. S3(a, b)], a smaller vertical polarization $P_z$, and lower piezoelectric coefficients [Table~\ref{table1}]. 
The larger band gaps in InO$X$ ($X$ = Cl, Br) relative to InOI originate from shallower halogen $p$ orbitals (lowering the VBM) and stronger In-$s$/$X$-$p_z$ $\sigma$-hybridization from shorter In-$X$ bonds (raising the CBM). Additionally, reduced band splitting, caused by diminished SOC effects due to the lighter halogen mass and a smaller built-in electric field, also contributes.
Using a modified deformation potential theory,~\cite{lang2016} we found that InO$X$ ($X$ = Cl, Br) MLs exhibit high and nearly isotropic electron mobility ($\mu_e > 10^3$ cm$^2$/V/s) at room temperature, due to their small electron effective mass and deformation potential constants [Table S1]. In contrast, their hole mobility is two orders of magnitude lower than the electron mobility. The asymmetric carrier transport and intrinsic vertical polarization effectively suppresses photo-generated carrier recombination, which is an advantage for optoelectronic device applications.    

Finally, to justify the hexagonal structure of InO$X$ ($X$ = Cl, Br, I) MLs, we performed comprehensive swarm-intelligence-based structural searches using the CALYPSO code.\cite{Wang2012} Our structural search revealed that the ground state of InO$X$ MLs is a rectangular monolayer ($Pmmn$ symmetry) [Fig. S10], exfoliated from the bulk layered lattice.\cite{Tasawar2025} The proposed hexagonal phases of InO$X$ ML are metastable, with energies 0.34--0.81 eV/f.u. above the ground state. Such metastability is experimentally achievable, as demonstrated by the successful synthesis of metastable 1T or 1T$^{\prime}$ phases of \ce{MoS2} ML,~\cite{Jayabal2018,Liu2018} which have energies 0.55--0.84 eV/f.u above the ground 1H phase.~\cite{Duerloo2014} 
Hexagonal InO$X$ ($X$ = Cl, Br, I) MLs can be obtained from the aforementioned decomposition process of the presynthesized InO crystal, without requiring a phase transition from the ground state by overcoming a high energy barrier. Furthermore, MD simulations confirm the kinetic stability of hexagonal InO$X$ ($X$ = Cl, Br, I) MLs, as no phase transitions between different crystal systems occur over extended trajectories under ambient conditions [Fig. S11]. 

In conclusion, we systematically investigated the structural phase transition and phase-dependent electronic/optical properties of hexagonal InOI ML using first-principles calculations. We identified two stable phases: a metallic T$^{\prime}$ phase and a topological insulating T phase. The T$^{\prime}$ phase is energetically favorable over the T phase across a wide strain range (-5\% to 2.5\%). The T$^{\prime}$-to-T phase transition, proceeding through the iodine atom diffusion, exhibits a moderate energy barrier of 72.1 meV/f.u.. This transition induces a metal-to-TI transition under compressive to neutral strains (-5\% to 0.5\%), while switching to a metal-to-normal insulator transition at tensile strains ($>0.5$\%). Moderate compressive strains ($-3$\% to 0\%) lower the energy barriers and enhance the topological properties of T-InOI ML simultaneously. Moreover, this phase transition shifts the optical absorption from the infrared to the visible zone and induces strong piezoelectricity in the T phase, with coefficients $d_{22} = 22.48$ pm/V and $d_{32} = 0.35$ pm/V. These multifunctional properties demonstrate the potential of InOI ML for applications in non-volatile memory, low-contact-resistance spintronics, and optical switching devices.\newline

See supplemental material for the potential energy surface, phonon dispersion, electronic structure, structural search, charge transfer, and the MD simulations.
%\newline

\begin{acknowledgments}
We acknowledge Dr. Yan Gao and Dr. Yunyun Bai inYanshan University for their fruitful discussions. This work was supported by the National Natural Science Foundation of China (No. 11904313), Cultivation Project for Basic Research and Innovation of Yanshan University (No.2022LGZD001), and the Innovation Capability Improvement Project of Hebei province (No. 22567605H). 
\end{acknowledgments}

\section*{Data Availability Statement}
The data that support the findings of this study are available from the corresponding author upon reasonable request.

\nocite{*}
\bibliography{inoi}% Produces the bibliography via BibTeX.
\end{document}